\documentclass[twocolumn,aps,showpacs,prx]{revtex4-1}

\usepackage{amsmath,amssymb,graphics,epsfig,epstopdf,color,multirow,array,verbatim,ulem,braket,tabularx}
\usepackage[colorlinks,linkcolor=blue,citecolor=blue,urlcolor=blue]{hyperref}
\usepackage{color}
\usepackage{graphicx}
\usepackage{epstopdf}
\usepackage{amsmath}
\usepackage{multirow}
\usepackage{ulem}
\usepackage{orcidlink}

\begin{document}

\title{Origin of insulating-like behavior of Bi$_2$Sr$_2$CaCu$_2$O$_{8+x}$ under pressure: A first-principles study}

\author{Xin Du\orcidlink{0000-0003-1918-7568}$^{1,2}$}
\author{Jian-Feng Zhang\orcidlink{0000-0001-7922-0839}$^{3}$}
\author{Zhong-Yi Lu\orcidlink{0000-0001-8866-3180}$^{1,2}$}\email{zlu@ruc.edu.cn}
\author{Kai Liu\orcidlink{0000-0001-6216-333X}$^{1,2}$}\email{kliu@ruc.edu.cn}

\affiliation{$^1$School of Physics and Beijing Key Laboratory of Opto-electronic Functional Materials $\&$ Micro-nano Devices, Renmin University of China, Beijing 100872, China \\
$^2$Key Laboratory of Quantum State Construction and Manipulation (Ministry of Education), Renmin University of China, Beijing 100872, China \\
$^3$Center for High Pressure Science $\&$ Technology Advanced Research, Beijing 100094, China}

\date{\today}

\begin{abstract}
Recent experimental study on Bi$_2$Sr$_2$CaCu$_2$O$_{8+x}$ superconductors has revealed an unexpected quantum phase transition from superconducting state to insulatinglike state under pressure [Zhou \textit{et al.}, Nat. Phys. 18, 406 (2022)]. To better understand the physical origin of this pressure-induced phenomenon, here we have studied the structural, electronic, and magnetic properties of undoped and O-doped Bi$_2$Sr$_2$CaCu$_2$O$_{8+x}$ (Bi2212) under pressures based on density-functional theory calculations. We first identified the crystal structure of undoped Bi2212 with the armchair distortions in the BiO layers and reproduced the insulating feature of the parent compound. Then we added an extra O atom to the parent compound to simulate the hole-doping effect and found that the structure with O dopant located in the van der Waals (vdW) gap is energetically the most stable. Further calculations on O-doped (0.125 holes/Cu) Bi2212 revealed that the pressure can induce charge redistributions between CuO$_2$ planes and BiO layers; specifically, Cu-$d_{x^2-y^2}$ orbitals gain electrons and Cu atoms rather than O atoms dominate around the Fermi level under high pressure. Along with the increasing pressure, the density of states at the Fermi level first reaches the maximum at $\sim$ 10 GPa and then shows a valley near the Fermi level above 20 GPa, which may be responsible for the insulatinglike state observed in recent experiments. We suggest that the competition among several factors, such as the increase of electrons in the CuO$_2$ plane, the variation of in-plane hopping due to the shortened Cu-O distance, and the enhanced Coulomb repulsion among the Cu-3$d$ electrons, could lead to the exotic transition under pressure. Our work provides an explanation of the high-pressure behaviors of Bi2212, which may facilitate a comprehensive understanding of cuprate superconductors.
\end{abstract}


\maketitle

\section{Introduction}
Due to the high superconducting transition temperature beyond McMillan limit at ambient pressure, cuprate superconductors have been a focal point in condensed-matter physics since their discovery decades ago \cite{Marezio93,Zaanen15,Shen03,Plakida10,Xue21,Hwang19,Bollinger16,Zaanen06,Abbamonte22}. Nevertheless, the strong correlation effect, the coupling among multiple degrees of freedom (spin, charge, orbital, and lattice), and the rich types of competing orders have made their superconducting mechanism elusive \cite{8,9,11,Lee06,10,He21,Liu24}. To find the key factors, different control methods, such as chemical doping and pressure, have been adopted to effectively modify the properties of cuprate superconductors, based on which the phase diagram of superconducting transition temperature ($T_\text{c}$) with the tuning parameter has been established \cite{Zaanen15,Shen21,Sun23}. According to the phase diagram of cuprates, extensive theoretical and experimental studies have been devoted to explore the origin of superconductivity from the normal states \cite{Wen06,Damascelli05,Hussey08,Shen14}. Despite several possible superconducting mechanisms including the preformed pairs \cite{Larkin97} or the spin fluctuations \cite{10} having been proposed, the intricate correlation between the normal state and the superconducting state still remains unclear.

Among multitudinous cuprate superconductors, the Bi-based cuprates stand out due to their high $T_\text{c}$ and high-quality monolayer crystals, which can serve as an ideal system for exploring the unconventional superconductivity \cite{Zhang19,Chen23,Zhang13,Davis02,Davis22}. In addition, the layered structure of Bi-based cuprates with van der Waals gap also renders them extremely susceptible to the external pressure. Recent experiments have systematically investigated the behaviors of Bi$_2$Sr$_2$CaCu$_2$O$_{8+x}$ (Bi2212) superconductors at different doping levels under pressure ($<$ 50 GPa) \cite{Long22}. Surprisingly, along with the increasing pressure, the superconducting $T_\text{c}$ of Bi2212 first increases and then decreases until the superconductivity completely disappears, and finally the system transforms into an insulatinglike state. The exotic experimental phenomena challenge the conventional wisdom that pressure often increases the bandwidth and favors the metallic properties. In fact, many other cuprate superconductors (such as YBa$_2$Cu$_3$O$_{7-}$$_\delta$ \cite{Schilling91,Chu11} and Tl$_2$Ba$_2$CaCu$_2$O$_{8+}$$_\delta$ \cite{Chen15}) also exhibit the suppressed superconductivity under pressure, indicating that the pressurized superconducting to insulatinglike state transition is quite common in the hole-doped cuprate superconductors \cite{Sun23}.

To reveal the physical mechanism of the pressure-induced transition between superconducting and insulatinglike states, taking Bi2212 as an example, we performed detailed first-principles calculations under pressure. Based on the stable O-doped structure, we studied the electronic and magnetic properties at different pressures and found that the pressure-induced lattice compression can reduce the energy differences among antiferromagnetic (AFM) states and cause the charge redistributions between Cu and O atoms in the CuO$_2$ plane. Interestingly, the evolution of the density of states at the Fermi level with pressure is similar to the dome-shape phase diagram of $T_\text{c}$ vs pressure in experiment. Moreover, the unusual valley in the density of states near the Fermi level above 20 GPa may be responsible for the appearance of insulatinglike state at high pressures. Our work provides helpful clues for understanding the emergent phenomena in cuprate superconductors under pressure.

\section{Computational details}
The calculations of structural, electronic, and magnetic properties of Bi2212 were carried out within the framework of fully spin-polarized density-functional theory (DFT) \cite{Kohn64,Sham65} as implemented in the Vienna \textit{ab initio} simulation package (VASP) \cite{Kresse96}. The projector augmented-wave (PAW) \cite{Blochl94} potentials with valence electron configurations of 5$d^{10}$6$s^2$6$p^3$, 4$s^2$4$p^6$5$s^2$, 3$s^2$3$p^6$4$s^2$, 3$d^{10}$4$s^1$, and 2$s^2$2$p^4$ were adopted for Bi, Sr, Ca, Cu, and O atoms, respectively. The generalized gradient approximation (GGA) of Perdew-Burke-Ernzerhof (PBE) type \cite{Ernzerhof96} was chosen for the exchange-correlation functional. To achieve good convergence, the plane-wave cutoff energy was set to 550 eV. The 12$\times$12$\times$2 and 8$\times$8$\times$2 Monkhorst-Pack {\bf k} meshes \cite{34} were used for the $\sqrt{2}$$\times$$\sqrt{2}$$\times$1 and 2$\times$2$\times$1 supercells in all calculations, respectively. The convergence thresholds were set as 1 × 10$^{-6}$ eV for energy and 0.01 eV/{\AA} for force. The Gaussian smearing method for Fermi surface broadening with a width of 0.05 eV was adopted for the local density of states (LDOS) calculations. To better describe the strong correlation effect among Cu 3$d$ electrons in Bi2212, the Hubbard interaction $U$ with the effective value of 9.13 eV based on the linear-response theory \cite{35} was adopted. The calculated lattice parameters ($a$ = $b$ = 3.740 {\AA} and $c$ = 31.041 {\AA}) of Bi2212 in antiferromagnetic ground state of $I$4/$mmm$ structure under ambient pressure are in good accordance with the experimental values ($a$ = $b$ = 3.814 {\AA} and $c$ = 30.52 {\AA}) \cite{Hwang88}. Notably, while the linear-response method for estimating $U$ may not fully capture all physical properties, different $U$-determination methods (e.g., unrestricted Hartree-Fock or linear-response method) obtained consistent $U$ values ($\Delta$$U$ $<$ 0.9 eV) for open-shell systems such as NiO and FeO \cite{Carter14}. This suggests that for open-shell cuprates like Bi2212 with Cu 3$d$$^9$ configuration, the choice of calculation method may have limited impact on the determined $U$ value. Furthermore, our calculations confirm that adopting different $U$ values (2.7, 4.7, or 8.0 eV) \cite{Moritz17,Bansil20} hardly changes the lattice parameters and the antiferromagnetic ground state under ambient pressure. Even under pressure, slight $U$-value variations
(e.g., comparing $U$ = 9.13 eV with $U$ = 8.0 eV) would not change the magnetic and electronic properties, verifying the validity of our calculations. Moreover, the DFT-D2 method \cite{Scoles01,Grimme06} was included to account for the van der Waals interactions between different BSCCO layers.

\begin{figure*}[!t]
	\includegraphics[angle=0,scale=0.5]{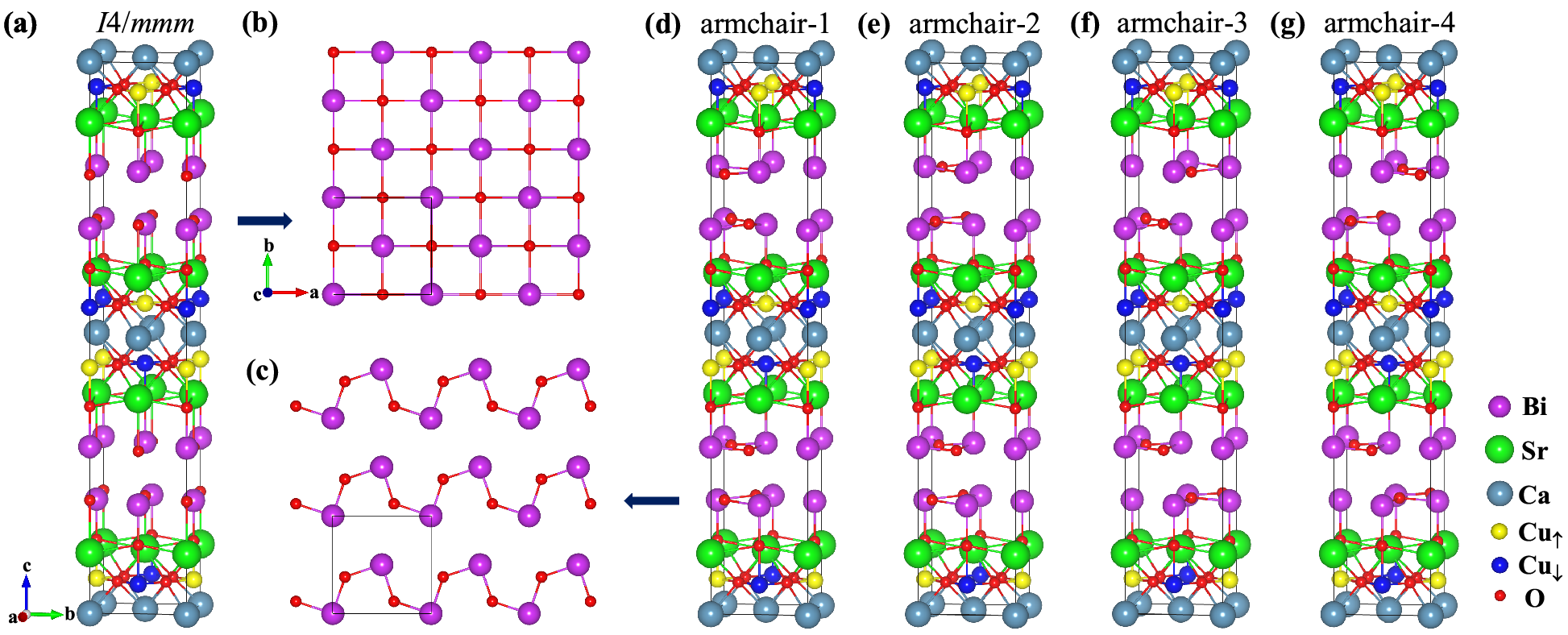}
	\caption{(Color online) (a) High-symmetry and (d)-(g) low-symmetry crystal structures of undoped Bi2212 under ambient pressure. Top views of the BiO layers in (b) high-symmetry and (c) low-symmetry structures from the $c$ axis, respectively. The carmine, green, grey-blue, yellow, blue, and red balls represent Bi, Sr, Ca, spin-up Cu, spin-down Cu, and O atoms, respectively.}
	\label{fig1}
\end{figure*}

\section{Results and discussion}

\subsection{Crystal structure and magnetic configuration}
As shown in Fig. \ref{fig1}(a), the undoped Bi$_2$Sr$_2$CaCu$_2$O$_8$ has a tetragonal structure with the $I$4/$mmm$ symmetry. Each repeating unit contains two CuO$_2$ planes that are separated by one Ca layer and are sandwiched between two SrO and BiO layers. In the CuO$_2$ plane, a Cu atom is connected with four O atoms, forming a two-dimensional (2D) square lattice. By using first-principles calculations, we firstly confirmed that Bi$_2$Sr$_2$CaCu$_2$O$_8$ adopts the AFM N\'{e}el ground state with a local magnetic moment of 0.6 $\mu$${\rm _B}$ on Cu. Further local density of states (LDOS) analyses show that it exhibits metallic characteristics due to the orbital contributions from BiO layers around the Fermi level (Fig. S1(a) of Supplemental Material (SM) \cite{SI}). These calculations on the crystal structure and electronic properties of undoped Bi$_2$Sr$_2$CaCu$_2$O$_8$ are in good agreement with the results of undistorted Bi2212 in previous studies \cite{Moritz17,Bansil20,Beigi24,Bansil06}.

It has been reported that in Bi2212 there exists structural supermodulation \cite{Bansil20,Beigi24,Zeng11,Yu94,Barboux89} with the armchairlike distortion in the BiO layers [Fig. \ref{fig1}(c)], which not only lowers the energy compared with the tetragonal phase [Fig. \ref{fig1}(b)] but also exhibits the insulatinglike property (Figs. S1(b) and S1(c) of SM \cite{SI}). Considering the relative orientations of the armchair BiO chains in different BiO layers, we built four structures (armchair-1, armchair-2, armchair-3, and armchair-4) based on the $\sqrt{2}$$\times$$\sqrt{2}$$\times$1 supercell of the $I$4/$mmm$ phase in the AFM N\'{e}el state [Figs. \ref{fig1}(d)-\ref{fig1}(g)]. In the armchair-1 and armchair-3 (armchair-2 and armchair-4) structures, the BiO armchair chains in the top and bottom BiO layers of each BSCCO monolayer are oriented in the same (opposite) directions. However, the BiO armchair chains in the adjacent BiO layers of different BSCCO monolayers are aligned in the armchair-1 or armchair-2 structures but misaligned in the other two. The total energy calculations show that the distorted structures are energetically $>$ 0.4 eV/Cu lower than the undistorted structure. Moreover, the armchair-2 structure has subtle energy difference with the armchair-3 and armchair-4 structures due to the weak interlayer interaction across the vdW gap, and the armchair-1 structure tends to relax to the NM state (Table S1 of SM \cite{SI}). The calculated LDOS of armchair-2 and armchair-4 structures of undoped Bi2212 well reproduce the insulatinglike feature, which can remove the contribution of the BiO layers at the Fermi level compared with that of the high-symmetry undistorted structure (Fig. S1 of SM \cite{SI}). It should be noted that when we considered the pressure effect, the energy of armchair-4 structure is slightly lower (0.006 eV/Cu) than that of amchair-2 structure at 45 GPa.

\begin{figure}[!b]
	\includegraphics[angle=0,scale=0.6]{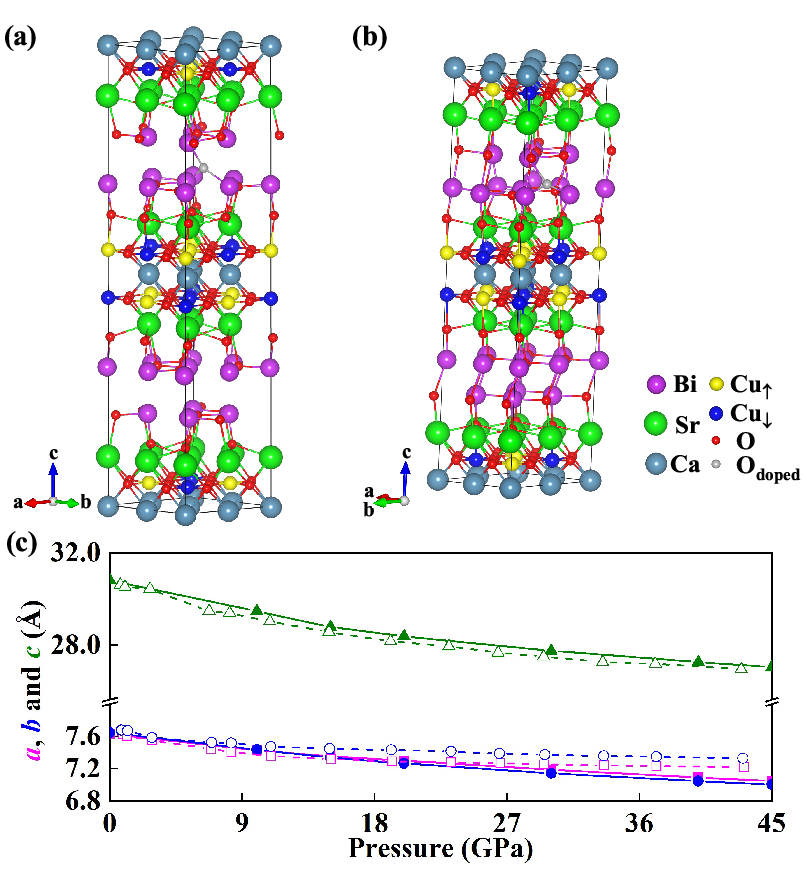}
	\caption{(Color online) Optimized crystal structure of the O-doped (0.125 holes/Cu) Bi2212 under (a) ambient pressure and (b) 45 GPa. The carmine, green, grey-blue, yellow, blue, red, and gray balls represent Bi, Sr, Ca, spin-up Cu, spin-down Cu, O, and extra doped O atoms, respectively. (c) Pressure dependence of lattice parameters ($a$, $b$, and $c$). The solid and dashed lines represent our calculated values and previous experimental values \cite{Long22,Hemley23}, respectively.}
	\label{fig2}
\end{figure}

Based on the $\sqrt{2}$$\times$$\sqrt{2}$$\times$1 supercell of armchair-4 structure, we added an extra O atom into the supercell to simulate the hole-doping effect (0.125 holes/Cu). Previous scanning-tunneling microscopy (STM) and theoretical works have proposed two possible sites for O dopant \cite{Bansil20,Hoffman12,Lin14}: one is between the BiO and SrO layers, located directly above the O atom in the CuO$_2$ plane (Figs. S2(a) and S2(d) of SM \cite{SI}); the other is in the BiO layer (Figs. S2(b) and S2(e) of SM \cite{SI}). Besides above two sites, we also considered the position in the vdW gap between the adjacent BiO layers, denoted as Type-C site (Figs. S2(c) and S2(f) of SM \cite{SI}). Then, we compared the total energies for those structures of O-doped Bi2212 in the nonmagnetic (NM) state under 0 GPa and 45 GPa (Table S2 of SM \cite{SI}). Surprisingly, no matter where the initial position of the doped O atom is, after the structural optimization all structures relax to Type-C-like structures with O dopant located between the two BiO layers (Fig. \ref{fig2}). In addition, we considered the NM state, the ferromagnetic (FM) state, and two AFM states (N\'{e}el and stripe) to study the magnetic ground state of the O-doped Bi2212. As shown in Table S3 of SM \cite{SI}, the AFM N\'{e}el state is the most stable one under ambient pressure or 45 GPa, while the energy differences among AFM states decrease with pressure. Moreover, we find that the crystal structure under pressure is obviously compressed, where the vdW gaps disappear and the adjacent BiO layers are bonded together [Figs. \ref{fig2}(a) and \ref{fig2}(b)]. We then carefully explored the evolution of the lattice parameters in the pressure range from 0 to 45 GPa. Apparently, the lattice parameter $c$ changes faster than those along $a$ and $b$ axes [Fig. \ref{fig2}(c)], which is consistent with the experimental observations \cite{Long22,Hemley23}.

\subsection{Electronic structure}
\begin{figure}[!t]
	\includegraphics[angle=0,scale=0.56]{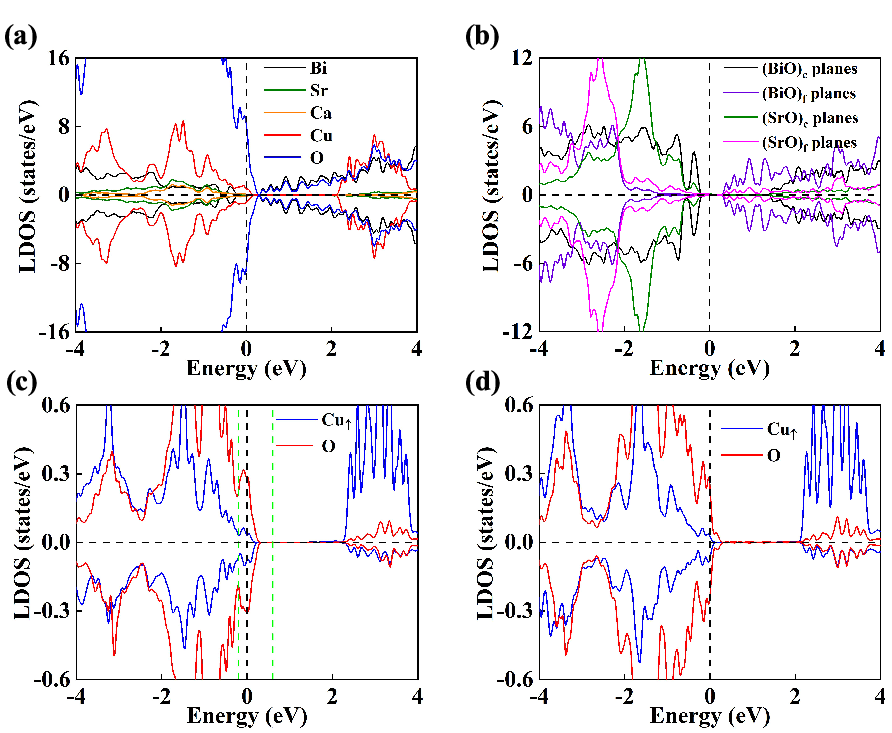}
	\caption{(Color online) (a) Total, (b) BiO-layer and SrO-layer, and (c) and (d) CuO$_2$-plane LDOS of the O-doped (0.125 holes/Cu) Bi2212 under ambient pressure. Here, the four BiO/SrO planes are classified into two types: close to and far away from the doped O, denoted as (BiO/SrO)$_\text{c}$ planes and (BiO/SrO)$_\text{f}$ planes. (c) and (d) Only one spin-up Cu and one O atom in the (CuO$_2$)$_\text{c}$ and (CuO$_2$)$_\text{f}$ planes are shown, respectively. The area between green lines in (c) represents the energy range of the partial charge densities plotted in Fig. \ref{fig5}.}
	\label{fig3}
\end{figure}
\begin{figure}[h]
	\includegraphics[angle=0,scale=0.564]{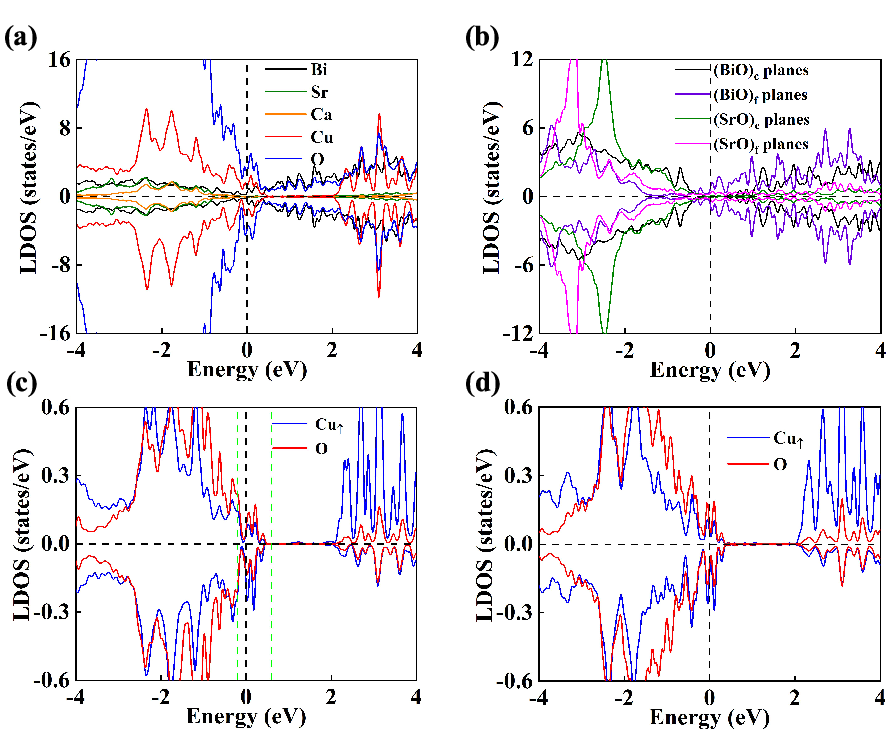}
	\caption{(Color online) (a) Total, (b) BiO-layer and SrO-layer, and (c) and (d) CuO$_2$-plane LDOS of the O-doped (0.125 holes/Cu) Bi2212 at 45 GPa. Here, the four BiO/SrO planes are classified into two types: close to and far away from the doped O, denoted as (BiO/SrO)$_\text{c}$ planes and (BiO/SrO)$_\text{f}$ planes. (c) and (d) Only one spin-up Cu and one O atom in the (CuO$_2$)$_\text{c}$ and (CuO$_2$)$_\text{f}$ planes are shown, respectively. The area between green lines in (c) represents the energy range of the partial charge densities plotted in Fig. \ref{fig5}.}
	\label{fig4}
\end{figure}
Pressure-induced structural distortions are usually associated with interesting changes of electronic and magnetic properties. Hence, we performed systematic calculations on the electronic structures of Bi2212 to explore the pressure effect.

\begin{table*}[!t]
	\caption{\label{table1} The integral values of the LDOS below the Fermi level for Cu5 and O40 in the (CuO$_2$)$_\text{c}$ plane of the O-doped Bi2212 under 0 GPa and 45 GPa.}
	\begin{center}
		\begin{tabular*}{18cm}{@{\extracolsep{\fill}} ccccccccc}
			\hline \hline
			Pressure& Cu5-$d_{xy}$& Cu5-$d_{yz}$& Cu5-$d_{z^2}$&  Cu5-$d_{xz}$& Cu5-$d_{x^2-y^2}$&  O40-$p_{y}$& O40-$p_{z}$&  O40-$p_{x}$\\
			\hline
			0 GPa& 1.982& 1.960& 1.935&  1.961& 1.378&  1.195& 1.191&  1.074\\
			45 GPa& 2.007& 1.974& 1.944&  1.971& 1.534&  1.272& 1.230&  1.084\\
			\hline \hline
		\end{tabular*}
	\end{center}
\end{table*}

First, we obtained the layer-resolved LDOS of O-doped Bi2212 under ambient pressure (Fig. \ref{fig3}). The total LDOS shows metallic properties and the states around the Fermi level are mainly contributed by O and Cu atoms [Fig. \ref{fig3}(a)]. In comparison, the DOSs of all BiO and SrO layers exhibit insulating behavior [Fig. \ref{fig3}(b)], implying the metallicity is derived from the CuO$_2$ planes. To simplify the analysis of the LDOS of CuO$_2$ plane, we only show the DOSs of one spin-up Cu atom and one O atom in Figs. \ref{fig3}(c) and \ref{fig3}(d). Like typical cuprates, the CuO$_2$ plane of undoped Bi2212 has a charge-transfer gap ($\sim$ 2 eV), where the conduction-band minimum and the valence-band maximum are mainly contributed by Cu and O orbitals, respectively. Once the hole doping is taken into account, the Fermi level shifts downward into the O orbitals of valence bands. It should be noted that here we studied two different CuO$_2$ planes, namely, the (CuO$_2$)$_\text{c}$ plane close to and the (CuO$_2$)$_\text{f}$ plane far away from the doped O atom between the adjacent BiO layers (Fig. \ref{fig2}). Although both of them show metallic properties, the doping effect of the former is stronger [Figs. \ref{fig3}(c) and \ref{fig3}(d)]. In real materials, the O dopants in experiment would have a random distribution. The properties of the (CuO$_2$)$_\text{c}$ plane close to O dopant are representative for doped Bi2212, and our following analyses also focus on this CuO$_2$ plane.

\begin{figure}[!b]
	\includegraphics[angle=0,scale=0.564]{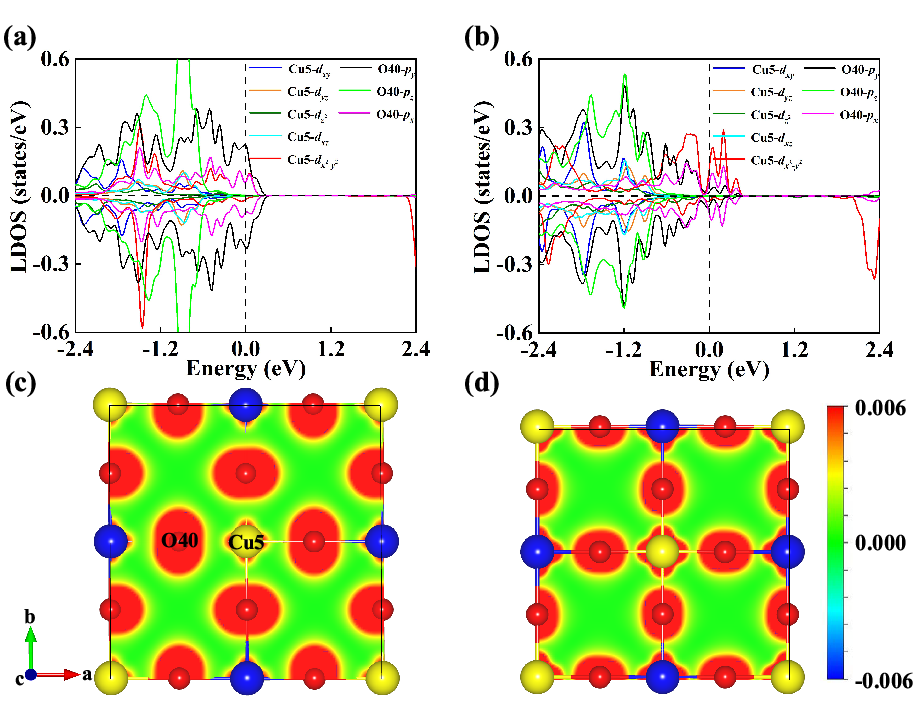}
	\caption{(Color online) (a) and (b) LDOS for Cu5 and O40 atoms in the (CuO$_2$)$_\text{c}$ plane of hole-doped (0.125 holes/Cu) Bi2212 under ambient pressure and at 45 GPa, respectively. (c) and (d) Partial charge-density maps of (CuO$_2$)$_\text{c}$ plane around the Fermi level [corresponding to the area between two green lines in Figs. \ref{fig3}(c) and \ref{fig4}(c)]. The color bar is in units of $e$/{\AA}$^3$.}
	\label{fig5}
\end{figure}

Then, we explored the electronic properties of the O-doped Bi2212 at 45 GPa. The total LDOS still exhibits metallic properties [Fig. \ref{fig4}(a)], while the states around Fermi level are significantly less than those of LDOS under ambient pressure [Fig. \ref{fig3}(a)]. Further layer-resolved DOS calculations show the (BiO)$_\text{f}$ and (SrO)$_\text{f}$ planes have weak metallic properties due to the compression of the crystal structure under pressure [purple and carmine lines in Fig. \ref{fig4}(b)], while the states around Fermi level are mainly attributed to CuO$_2$ planes [Figs. \ref{fig4}(c) and \ref{fig4}(d)]. Notably, compared with the dominant role of O orbitals around the Fermi level under ambient pressure [Fig. \ref{fig3}(c)], here at 45 GPa the Cu and O atoms in the (CuO$_2$)$_\text{c}$ plane have comparable contributions around the Fermi level, and an unusual valley in the DOS appears near the Fermi level ($\sim$ -0.07 eV) [Fig. \ref{fig4}(c)]. And the properties of the (CuO$_2$)$_\text{f}$ plane are quite similar to those of the (CuO$_2$)$_\text{c}$ plane, but show weaker doping effect [Fig. \ref{fig4}(d)]. Moreover, we examined the overdoped case (0.25 holes/Cu) by introducing two O atoms into Bi2212 supercell under both ambient pressure and 45 GPa (Figs. S3, S4, and Table S4 of SM \cite{SI}). The results for the overdoped case are well consistent with our previous findings for the near-optimal doping (0.125 holes/Cu), which not only confirms the validity of our previous calculations, but also aligns with experimental observations that both optimally doped and overdoped Bi2212 undergo a pressure-induced superconducting-to-insulatinglike transition \cite{Long22}.

\begin{figure}[!b]
	\includegraphics[angle=0,scale=0.56]{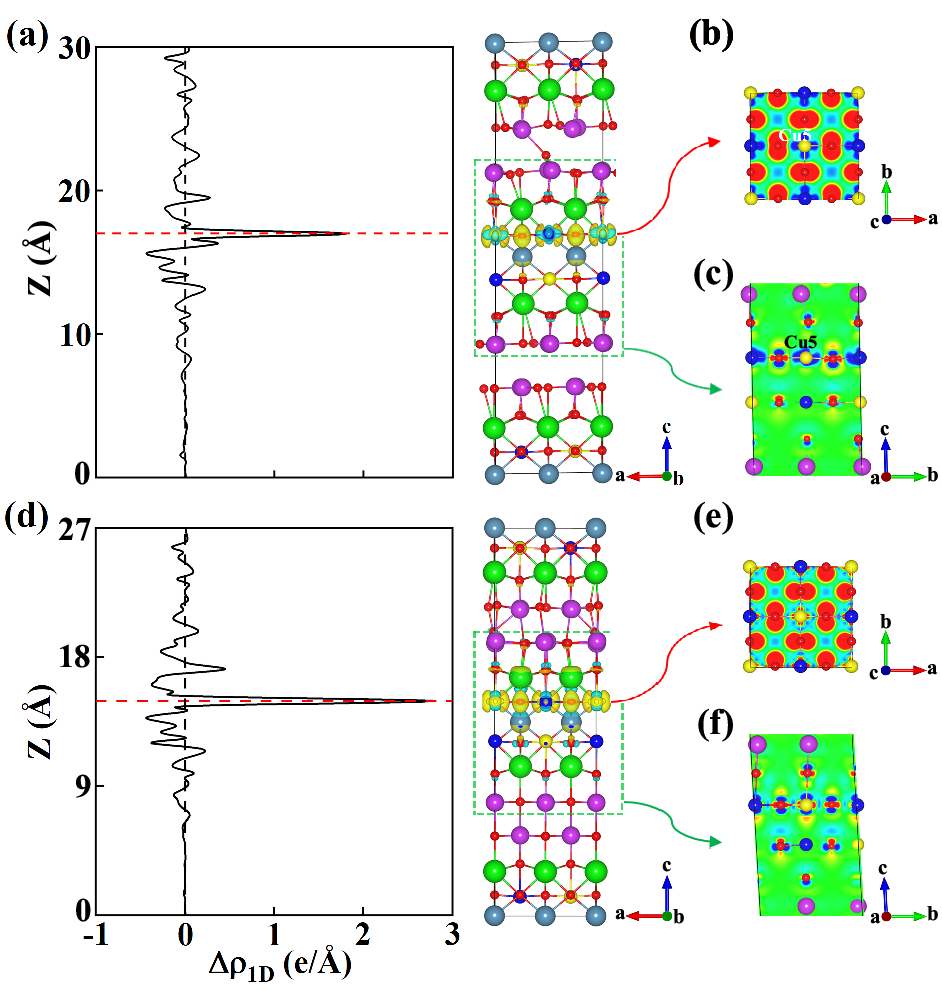}
	\caption{(Color online) 1D, 3D, and 2D differential charge densities for O-doped (0.125 holes/Cu) Bi2212 in the AFM N\'{e}el state under (a)-(c) ambient pressure and (d)-(f) at 45 GPa. (b) and (e) are the 2D differential charge densities in the (001) plane. (c) and (f) are the 2D differential charge densities in the (100) plane. The red dashed lines in (a) and (d) label the (CuO$_2$)$_\text{c}$ plane. The yellow and cyan isosurfaces in (a) and (d) represent the electron accumulation and depletion areas, respectively. The red and blue areas in (b), (c), (e), and (f) represent the electron gain and loss regions, respectively. The isosurface value is set to 0.006 $e$/{\AA}$^3$.}
	\label{fig6}
\end{figure}

Next, in order to analyze the origin of the changes in electronic properties under pressure, we carefully compared the orbital-resolved DOS of Cu and O atoms in the (CuO$_2$)$_\text{c}$ plane and the corresponding partial charge density around the Fermi level under ambient pressure and 45 GPa (Figs. \ref{fig5} and S5 of SM \cite{SI}). A notable feature in LDOSs is the change of orbitals around the Fermi level [Figs. \ref{fig5}(a) and \ref{fig5}(b)]. The states around the Fermi level at ambient pressure are mainly from O-$p_{y}$ and O-$p_{x}$ orbitals [Fig. \ref{fig5}(a)], while at 45 GPa the contribution of Cu-$d_{x^2-y^2}$ orbital is unexpectedly larger than the contribution of O orbitals (Cu-$d_{x^2-y^2}$ $>$ O-$p_{x}$ $>$ O-$p_{y}$) and their hybridization around the Fermi level is strongly enhanced [Fig. \ref{fig5}(b)]. To quantitatively analyze the charge variations, we integrated the values of the LDOS below the Fermi level for Cu5 and O40 atoms at 0 and 45 GPa (Table \ref{table1}), which intuitively shows that only the Cu-$d_{x^2-y^2}$ orbital significantly gains electrons under pressure. Partial charge-density maps [within the energy window of green lines in Figs. \ref{fig3}(c) and \ref{fig4}(c)] are shown in Figs. \ref{fig5}(c), \ref{fig5}(d), and S5 of SM \cite{SI}, which directly demonstrate the real-space charge distribution around the Fermi level. Under pressure, the amount of charge around the Cu5 atom increases obviously, and the change of the charge contours around the O atom corresponds to the variation of $p$ orbitals, which are in accordance with the DOS results of Figs. \ref{fig5}(a) and \ref{fig5}(b). Another notable feature is the unusual valley of the DOS in the vicinity of the Fermi level [Fig. \ref{fig5}(b)], which will be discussed in the subsequent section.

\begin{figure*}[!t]
	\includegraphics[angle=0,scale=0.75]{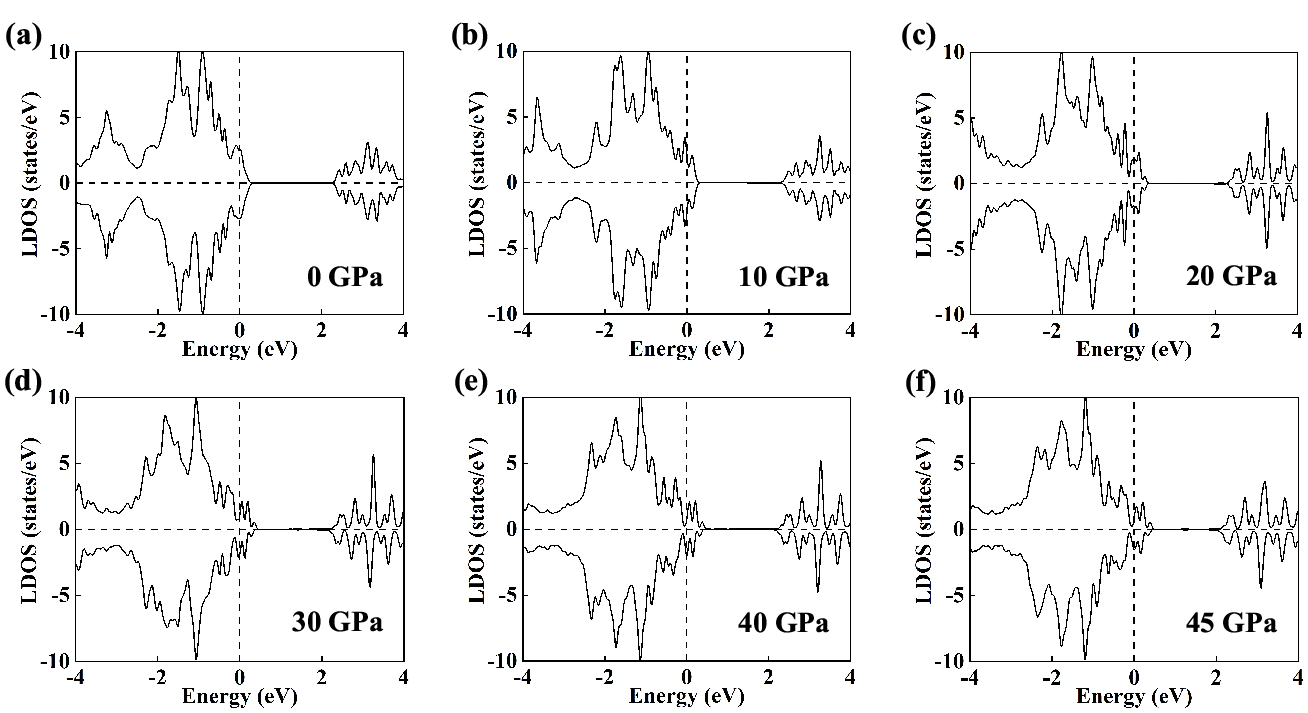}
	\caption{(Color online) The total LDOS of (CuO$_2$)$_\text{c}$ plane of O-doped (0.125 holes/Cu) Bi2212 under the pressures of (a) 0, (b) 10, (c) 20, (d) 30, (e) 40, and (f) 45 GPa, respectively.}
	\label{fig7}
\end{figure*}

Finally, to better understand the charge transfer in the O-doped Bi2212, we further calculated the one-dimensional (1D), two-dimensional (2D), and three-dimensional (3D) differential charge densities under ambient pressure and 45 GPa (Fig. \ref{fig6}), which can both qualitatively and quantitatively display the real-space charge variations. Specifically, the 3D differential charge-density map was obtained by subtracting the sum of the charge density of one isolated CuO$_2$ plane and that of the remaining structural components (excluding one CuO$_2$ plane) from the total charge density of the whole system. Furthermore, by integrating the 3D differential charge densities along the $x$ and $y$ directions, we obtained the 1D differential charge densities, which reflects the variation of the differential charge densities along the $z$ direction. As shown in Figs. \ref{fig6}(a) and \ref{fig6}(d), the 1D and 3D differential charge densities intuitively show that the (CuO$_2$)$_\text{c}$ plane obtains more electrons at 45 GPa than that under ambient pressure. This can be well explained from a structural perspective: since the O dopant can obtain more electrons from the upper BiO layer (Fig. \ref{fig2}) under pressure due to the shortened distance than that at ambient pressure, the bottom BiO layer can give more electrons to the (CuO$_2$)$_\text{c}$ plane. Actually, from the 2D differential charge densities, we can know more detailed information about the atomic orbitals. The top and side views of the (CuO$_2$)$_\text{c}$ plane under different pressures reflect that the transition from electron loss to electron gain in the Cu5-$d_{x^2-y^2}$ orbital [Figs. \ref{fig6}(b) and \ref{fig6}(e)] as well as the redistribution of electrons in the Cu5-$d_{z^2}$ orbital along the $c$ axis [Figs. \ref{fig6}(c) and \ref{fig6}(f)]. Overall, the above results of electronic properties show that the compression of the crystal structure due to the pressure effect can cause the charge redistributions between Cu and O atoms, which may affect the macroscopic properties such as superconductivity and transport behavior.

\section{Comparison with experiments}
Recent experiment has revealed that the pressurized Bi2212 undergoes an exotic transition from superconducting to insulatinglike state. Specifically, with the increasing pressure, the superconducting $T_\text{c}$ of optimally doped Bi2212 first increases ($T_\text{c-max}$ = 113 K at 10 GPa), then decreases, and then completely disappears at 37 GPa, above which the system shows an insulatinglike behavior up to 45 GPa \cite{Long22}. These results challenged the conventional understanding that the pressure usually increases the bandwidth and favors the metallic property. To better understand the experimental phenomenon, we studied the electronic properties of O-doped Bi2212 as a function of pressure. 

The total LDOS of (CuO$_2$)$_\text{c}$ plane in O-doped Bi2212 under different pressures are plotted in Fig. \ref{fig7}. Surprisingly, the density of states at the Fermi level first increases and then decreases with pressure, reaching the maximum at 10 GPa [Fig. \ref{fig7}(b)], which is similar to the dome-shape phase diagram of superconducting $T_\text{c}$ with pressure in experiment \cite{Long22}. In addition, we found that a DOS valley occurs near the Fermi level above 20 GPa [Figs. \ref{fig7}(d)-\ref{fig7}(f)], which may be responsible for the appearance of insulatinglike state under pressure.

Combined with all above calculations, we suggest that the observed property transition of Bi2212 under pressure \cite{Long22} may be the competition results of the following three factors: (i) The (CuO$_2$)$_\text{c}$ plane gains more electrons with increasing pressure (as shown in Fig. \ref{fig6} and Table \ref{table1}). Then the filled Cu-3$d$ orbitals and O-2$p$ orbitals in the (CuO$_2$)$_\text{c}$ plane may reduce the hole-type carriers and favor the insulating behavior. (ii) Under pressure, the in-plane lattice parameters are compressed (Fig. \ref{fig2}) and the orbital hybridization between Cu and O atom is strengthened (Figs. \ref{fig4} and \ref{fig5}), which increases the in-plane hopping, broadens the electronic bands, and reduces the local magnetic moments on Cu atoms. The increased in-plane hopping favors the metallic behavior. (iii) When the in-plane Cu-$d_{x^2-y^2}$ orbital gains enough electrons, the enhanced Coulomb repulsive interaction and the Cu-O orbital hybridization lead to band splitting in the lower Hubbard band (Figs. \ref{fig5} and \ref{fig7}), and as a result the (CuO$_2$)$_\text{c}$ plane tends to show insulatinglike behavior. In general, the mutual competition of these different factors induces the variation of superconductivities as well as the superconducting to insulatinglike state transition. We propose that at higher pressure the bands around the Fermi level may become broadening and the system may restore the  metallic properties again.

\section{Conclusion}
Based on spin-polarized density-functional theory calculations, we have investigated the structural, magnetic, and electronic properties of undoped and O-doped Bi$_2$Sr$_2$CaCu$_2$O$_8$ under pressure. First, we confirmed that the undoped Bi2212 in the AFM N\'{e}el state with the armchair structural distortion in BiO layers exhibits insulatinglike properties. Then we studied the O-doped Bi2212 and found that the Type-C structure with the O dopant locating in the van der Waals gap between the adjacent BiO layers of BSCCO units is energetically the most stable. Further comparisons of electronic properties under ambient pressure and 45 GPa show that the compression of the crystal structure due to pressure effect can cause the charge redistributions between Cu and O atoms. Especially, Cu-$d_{x^2-y^2}$ orbital obtains electrons and has a strong hybridization with O-$p$ orbitals around the Fermi level under pressure. Referring to the dome-shape phase diagram of the superconducting $T_\text{c}$ with pressure, our calculations show the density of states of CuO$_2$ plane at the Fermi level has the similar pressure dependence, which first increases and then decreases with pressure, reaching the maximum at 10 GPa. With further increasing pressure beyond 20 GPa, we suggest that the unusual valley in DOS near the Fermi level, that is, the secondary splitting of the lower Hubbard band, may be responsible for the appearance of insulatinglike state at high pressure. We ascribe the reason for the experimental observation of superconducting to insulatinglike state transition under pressure to be the result of competitions from several factors, such as the increase of 3$d$ electrons in the CuO$_2$ plane, the variation of in-plane hopping due to shortened Cu-O distance, and the enhanced Coulomb repulsive interaction among Cu-3$d$ electrons. Our findings provide an explanation for the puzzling phenomenon of the pressure-induced superconducting to insulatinglike state transition in Bi2212, which may facilitate the comprehensive understanding of the mystery of cuprate superconductors under pressure.

\begin{acknowledgments}
We thank Professor Liling Sun for helpful communication. This work was supported by the National Key R\&D Program of China (Grants No. 2022YFA1403103 and No. 2024YFA1408601) and the National Natural Science Foundation of China (Grants No. 12174443 and No. 12434009). Computational resources have been provided by the Physical Laboratory of High Performance Computing at Renmin University of China and the Beijing Super Cloud Computing Center.

\end{acknowledgments}

%\begin{appendix}

%\section{Phonon spectra}

%\section{Electronic band structure}

%\end{appendix}

\end{document}